 \documentclass[preprint,review,12pt]{elsarticle}

 \usepackage{graphicx}

\usepackage{amssymb}
\usepackage{epstopdf}

\journal{Physics Letters B}

\begin{document}

\begin{frontmatter}



\title{Predictions for net-proton and net-kaon distributions at LHC energies}


\author{Yacine Mehtar-Tani˜$^*$ and Georg Wolschin}

\address{$^*$Now at Departamento de F\'i˜sica de Part\'iculas and IGFAE, Universidade de Santiago de Compostela, Spain}
\address{ Institut f{\"ur} Theoretische 
Physik
der Universit{\"a}t Heidelberg, 
        Philosophenweg 16,  
        D-69120 Heidelberg, Germany}

\begin{abstract}
We investigate baryon and charge transport  in relativistic heavy-ion 
collisions, compare with Au + Au RHIC data at $\sqrt {s_{NN}}$ = 0.2 TeV, and make predictions for net-proton rapidity distributions
in central Pb + Pb collisions at CERN LHC energies of $\sqrt {s_{NN}}$ = 2.8,
3.9, and 5.5 TeV. We use the gluon saturation model and put
special emphasis on the
midrapidity valley  $|y|\leq 2$.
Net-kaon distributions are calculated and compared to BRAHMS Au + Au data at RHIC energies of
$\sqrt {s_{NN}}$ = 0.2 TeV, and predicted for Pb + Pb at 5.5 TeV.\\
\end{abstract}

\begin{keyword}
Relativistic heavy-ion collisions \sep Net-proton and net-kaon rapidity distributions \sep Saturation model
 \sep Predictions at LHC energies 
\PACS 24.85.+p \sep 25.75.-q \sep 25.75.Dw \sep 12.38.Mh

\end{keyword}

\end{frontmatter}





The test of gluon saturation in relativistic heavy-ion collisions is an important aim of the forthcoming Pb + Pb experiments at the LHC. At these energies gluons dominate the dynamical evolution of the system,
which is driven by a single hard scale, the saturation scale \(Q_s \gg \Lambda_{QCD}\) \cite{mcl94}.
Whereas most theoretical investigations concentrate on charged-hadron production from inclusive gluon interactions \cite{kha04,nar05}, the valence-quark scattering off the gluon condensate as an observable in net-baryon distributions \cite{mtw09} is expected to provide interesting new information on gluon saturation, and on geometric scaling \cite{sta01}.

Here the most promising effects arise at very forward angles, or correspondingly large values of the rapidity \(y \simeq 5 - 8\) at LHC energies of \(\sqrt {s_{NN}}\) = 5.5 TeV for Pb + Pb, with a beam rapidity of 8.68. For symmetric systems, two symmetric fragmentation peaks are expected to be present in the net-baryon distributions at forward/backward rapidities. In particular, we have shown in \cite{mtw09} that it is in principle possible to determine the growth of the saturation-scale exponent, 
\(\lambda \equiv d\ln{Q_s}/dy_b\), with the beam rapidity \(y_b\) from the position of the fragmentation peak in rapidity space. 

In the region of relatively large values of Feynman-\(x\simeq0.1\), the valence-quark parton distribution in the projectile is well-known close to and below its maximum, and can hence be used to access the gluon distribution at small \(x\) in the other nucleus where saturation is expected to occur due to the competition of gluon recombination with the exponentially increasing gluon splitting 
\cite{Bal96,Jal97,Ian01}.

Whereas it is interesting to investigate these effects theoretically, the forthcoming LHC experiments with heavy-ion capability ALICE, CMS, TOTEM and ATLAS initially will not be able to detect identified baryons and antibaryons from central heavy-ion collisions at large values of rapidity in the fragmentation-peak region. In particular, the dedicated LHC heavy-ion experiment ALICE will provide full particle identification for protons and antiprotons as well as kaons only in the central part of the net-baryon midrapidity valley \cite{aam08}.

As a consequence, one first has to concentrate on the midrapidity region in order to compare with data. In this Letter we present our predictions for net-baryon rapidity distributions in central Pb + Pb collisions at LHC energies. The LHC physics program starts with center-of-mass proton-proton energies of 7 and possibly 10 TeV, the corresponding energies for Pb + Pb (scaling with Z/A) are \(\sqrt {s_{NN}}\) = 2.76 and 3.94 TeV. We also present predictions for the highest attainable Pb + Pb energy of 5.52 TeV. Since experimental results will be available for net protons, we calculate these at the highest LHC energy in the midrapidity valley \(|y|<2\)
 instead of net baryons, and also include a prediction for net kaons 
\((K^+ - K^-)\) since these carry part of the valence quarks.


The differential cross section for valence quark production in a high-energy nucleus-nucleus collision is calculated from \cite{kha04}
\begin{equation} \label{eq:crossGS}
\frac{dN}{d^2p_Tdy}= \frac{1}{(2\pi)^2 } \frac{1}{ p_T^2}\;x_1q_v(x_{1},Q_{f})\;\varphi\left(x_2,p_T\right),
\end{equation}
where \(p_T\) is the transverse momentum of the produced quark, and \(y\) its rapidity. The longitudinal momentum fractions carried, respectively, by the valence quark in the projectile and the soft gluon in the target are \(x_1=p_T/\sqrt{s}\exp(y)\) and \(x_2=p_T/\sqrt{s}\exp(-y)\). The factorization scale is usually set equal to the transverse momentum, \(Q_{f}\equiv p_T\)
\cite{dum06}. 
We have discussed the gluon distribution \(\varphi(x,p_T)\) and details of the overall model in
 \cite{mtw09}.

The contribution of valence quarks in the other beam nucleus is added incoherently by changing \(y \to -y\). The valence quark distribution of a nucleus, \(q_v\equiv q-\bar q\), is given by the sum of valence quark distributions \(q_{v, N}\) of individual nucleons, \(q_v\equiv A q_{v, N}\), where \(A\) is the atomic mass number. 


Assuming that  the rapidity distribution for net baryons is proportional to the valence-quark rapidity distribution up to a constant factor \(C\), we obtain by integrating over \(p_T\),
\begin{equation} \label{eq:ydistquark}
\frac{dN}{dy}= \frac{C}{(2\pi)^2 }\int \frac{d^2p_T}{ p_T^2}\;x_1q_v(x_{1},Q_{f})\;\varphi\left(x_2,p_T\right).
\end{equation}
It turns out that this is indeed a good approximation at sufficiently high energy, in particular, when comparing to Au + Au data from RHIC, and we expect it to be valid at LHC as well
\cite{mtw09}.

The unintegrated gluon distribution is peaked at  \(q_T=Q_s\), or \(x_1=\exp\left(-\tau/2+\lambda\right)\), with the saturation momentum squared \(Q^2_s=A^{1/3} Q^2_0 x_2^{-\lambda}\), the saturation-scale exponent \(\lambda\), and the scaling variable \(\tau=\ln (s/Q_0^2) - \ln A^{1/3} - 2(1+\lambda)\,y\) that we have introduced in \cite{mtw09}. Here \(A\) is the nucleon number and \(Q_0\) sets the dimension. The peak at
\(q_T=Q_s\) reflects the fact that most of the gluons sit at this value. Therefore, we  expect \(dN/dy\sim x_1q(\langle x_1\rangle)\), with \(\langle x_1\rangle\equiv \langle Q_s\rangle/\sqrt{s} \exp(y)\). With \(x_2=x_1 \exp(-2y)\) we can solve this equation for \(\langle x_1 \rangle, \) yielding
\begin{equation}
\langle x_1 \rangle = \left(\frac{A^{1/6} Q_0}{\sqrt{s}}\right)^{1/(1+\frac{\lambda}{2})} \exp\left[2 \frac{1+\lambda}{2+\lambda} y\right]. \label{eq:meanx1}
\end{equation}

In the region of small \(x_1\) and \(x_2\)  corresponding to the midrapidity valley (\(y\sim 0\)) away from the peaks,  the valence quark distribution behaves as \(xq_v\propto x^\Delta\), where  the intercept 
\(\Delta\) has been calculated in the saturation picture  \cite{ita03} leading to
\begin{equation}
\Delta=1-\sqrt\frac{2\alpha_s C_F}{\pi(1-\lambda)}
\end{equation}
with \(C_F = (N_C^2-1)/ 2N_C\), \(N_C=3\). The value of \(\Delta\) had been fitted to the old preliminary BRAHMS data in \cite{ita03}, with \(\Delta\simeq 0.47\), leading to a strong-coupling constant 
\(\alpha_s\simeq 0.3\).

Therefore, in the midrapidity valley Eq. (\ref{eq:ydistquark}) becomes
\begin{equation}\label{eq:valley}
\frac{1}{A}\frac{dN}{dy} \propto \left(\frac{A^{1/6} Q_0}{ \sqrt{s} } \right)^{\Delta/(1+\frac{\lambda}{2})}\cosh\left[2\Delta \frac{1+\lambda}{2+\lambda} y\right]
\end{equation}
which reduces to Eq. (80) in \cite{ita03} for the special case \(\lambda = 0\).
The midrapidity values of the net-baryon or net-proton rapidity distributions at two different center-of-mass energies in the nucleon-nucleon system
are related through
\begin{equation}\label{eq:f}
\frac{dN}{dy}(s)= \left( \frac{s_{0}}{s} \right)^{\Delta/(2+\lambda)}
\frac{dN}{dy}(s_0).
\end{equation}


\begin{figure*}
\begin{center}%
\includegraphics[width=9cm]{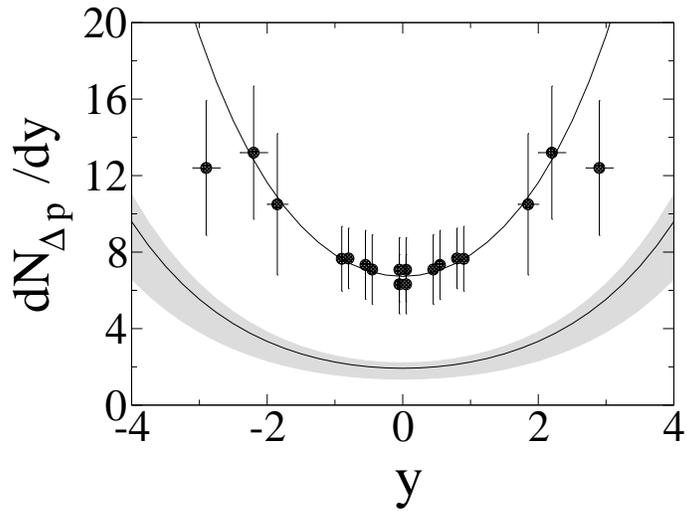}
\caption{\label{fig1} The rapidity distribution of net protons in central (0\%--5\%) Au + Au collisions at RHIC energies of \(\sqrt{ s_{NN}}\) = 0.2 TeV as measured by BRAHMS \cite{bea04} (black dots) is fitted with our theoretical formula using a \(\chi^2-\) minimization to fix the parameters for the predictions at LHC energies. The data point at y = 2.9 is neglected in the minimization. The grey band in the lower part of the figure shows our predictions for central Pb + Pb collisions at LHC energy of 
\(\sqrt {s_{NN}}\) =  2.76 TeV (corresponding to 7 TeV in  p + p) with \(\lambda = 0.3\) (upper bound), \(\lambda = 0.2\) (solid curve), and \(\lambda = 0\) (lower bound), using Eq. (\ref{eq:f}).}
\end{center}
\end{figure*}

\begin{figure}
\begin{center}
\includegraphics[width=9cm]{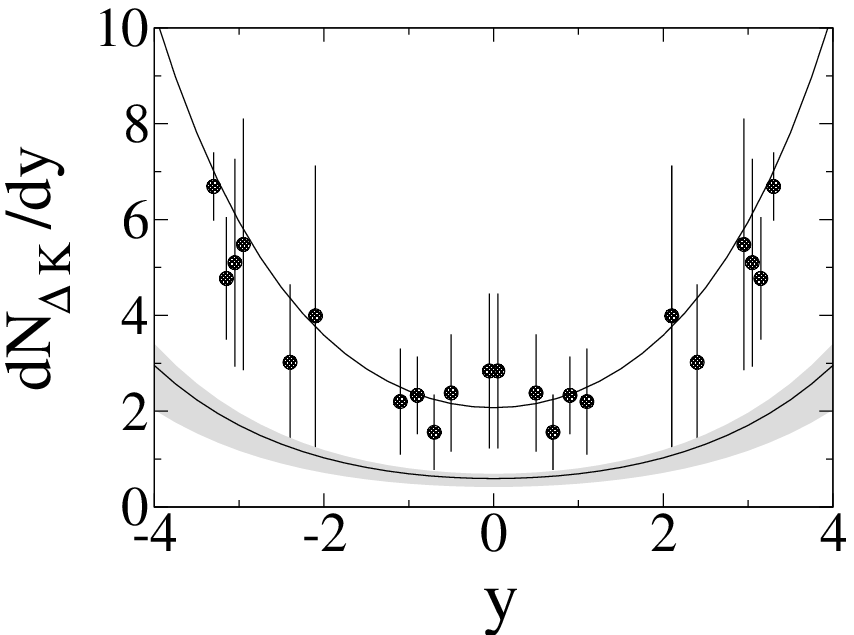}
\caption{\label{fig2}The net-kaon rapidity distribution in central (0\%--5\%) Au + Au collisions at RHIC energies of \(\sqrt{ s_{NN}}\) = 0.2 TeV as measured by BRAHMS \cite{bea04} (black dots) is fitted with our theoretical formula using a \(\chi^2-\) minimization. The grey band in the lower part of the figure shows our predictions for central Pb + Pb collisions at LHC energy of \(\sqrt {s_{NN}}\) =  2.76 TeV (corresponding to 7 TeV in p + p) with 
\(\lambda = 0.3\) (upper bound), 
\(\lambda = 0.2\) (solid curve), and \(\lambda = 0\) (lower bound).}
\end{center}
\end{figure}

We now use the analytical form for \(dN/dy = a \cosh(by)\) (cf. Eq. (\ref{eq:valley})) in a direct comparison with BRAHMS net-proton data in central 
Au + Au collisions at RHIC energies of \(\sqrt{ s_{NN}}\) = 0.2 TeV \cite{bea04} through a \(\chi^2-\) minimization of the two parameters \(a\) and \(b\), where \(b=2\Delta (1+\lambda)/(2+\lambda) \).

Our comparison with the BRAHMS Au + Au data in the midrapidity region is shown in
 Fig.~\ref{fig1} for net protons.
The fit parameters are \(a = 6.79\pm0.59\), \(b = 0.575 \pm 0.116\) as results of the \(\chi^2-\) minimization per degree of freedom (8 data points - 2 free parameters; \(\chi^2\)/dof = 0.028). With the energy dependence as expressed in Eq.(\ref{eq:f}), the grey band in the lower part of the figure shows our predictions for central Pb + Pb collisions at a LHC energy of \(\sqrt {s_{NN}}\) =  2.76 TeV with \(\lambda = 0.3\) (upper bound), \(\lambda = 0.2\) (solid curve), and \(\lambda = 0\) (lower bound). The
mass-number dependence is very weak, and we neglect it in the discussion (\(A_{Pb}/A_{Au} \simeq 1.056\)). 

Note that for \(\lambda = 0\), we have \(\Delta = b \). Our value for \(b\) is slightly larger than, but 
within our error bars compatible with the one fitted by \cite{ita03}. We extract a value of 
\(\Delta \approx 0.575 \) for \(\lambda = 0\), and \(\Delta \approx 0.509 \) for \(\lambda = 0.3\), leading to 
\(\alpha_s \simeq 0.2 \) in both cases due to compensating effects. 

Our result for the midrapidity distributions should be compared directly to the forthcoming ALICE net-proton data in central Pb + Pb collisions. The predicted midrapidity value at \(\sqrt {s_{NN}}\) =  2.76 TeV is \(dN/dy \simeq\) 1.93, it depends only slightly on the saturation-scale exponent \(\lambda\) and hence, one cannot expect to determine the value of  
\(\lambda\) from midrapidity net-baryon data. 

For a determination of \(\lambda\) from heavy-ion data at LHC energies, one therefore has to rely on the forward rapidity region for net baryons \cite{mtw09} which is difficult to access experimentally, or on midrapidity distributions for produced charged particles \cite{nar05}. From the overall accuracy of our prediction regarding the absolute value at midrapidity, and  the shape of the net-proton rapidity distribution, we will, however, be able to draw conclusions regarding the validity of the gluon saturation picture.

In the comparison of our model calculations with RHIC net-baryon data in \cite{mtw09},
medium and final-state effects turned out not to be important for rapidity distributions, although they may be visible in transverse momentum distributions. Due to baryon-number conservation, they could only lead to a redistribution in rapidity space, which is, however, not observed. This result differs qualitatively from the importance of medium effects in jet suppression at these energies.
The expectation is that net-baryon distributions at LHC are also not visibly affected by medium and final-state effects.

Like net baryons and net protons, net charged mesons such as kaons and pions carry part of the valence quarks, and can thus be treated on the same footing. In particular, these can be used as a cross-check for the validity of our hypothesis that net-baryon and net charged-meson rapidity distributions essentially reflect the valence quark distributions, such that hadronization does not play a significant role. Here we study the net-kaon rapidity distribution since we don't have access to the full net charged-meson distribution.

In Fig.~\ref{fig2} we show the result for the net-kaon rapidity distribution in  central 
Au + Au collisions at RHIC energies of \(\sqrt{ s_{NN}}\) = 0.2 TeV in comparison with BRAHMS 
data \cite{bea05} through a \(\chi^2-\) minimization of the two parameters, \(a = 2.087\pm 0.173\) and 
\(b = 0.535\pm 0.031\). Here the result of the minimization is \(\chi^2\)/dof = 0.540. Interestingly enough, the value of \(b\) for 
\(\Delta_K\) is compatible with the one extracted for net protons. This indicates that the rapidity distribution is primarily sensitive to the initial conditions of the collision, not to the hadronization process, since the slope is not depending on the species of the produced particles (protons or kaons). 

We had discussed the effect of fragmentation in our earlier work [4] by comparing calculations 
for net baryons with and without fragmentation function, cf. Fig.~5 in that work. In the midrapidity region
\( |y| < 2\), the effect at \(\sqrt{ s_{NN}}\) = 0.2 TeV is clearly smaller than the size of the experimental error bars. We have therefore not discussed the effect explicitly in the present work.
Again our extrapolation to central Pb + Pb at a LHC energy of 2.76 TeV
is shown by the band in the lower part of Fig.~2, with the solid curve for \(\lambda = 0.2\).
Our predicted midrapidity value is \(dN/dy \simeq\) 0.7.

\begin{figure}
\begin{center}
\includegraphics[width=9cm]{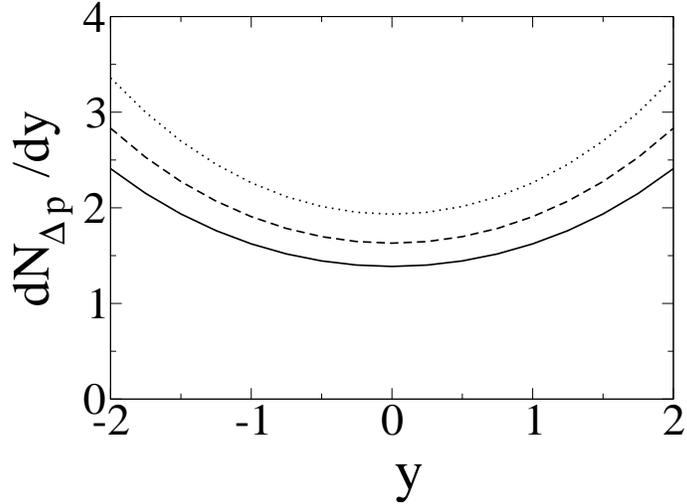}
\caption{\label{fig3}Rapidity distributions of net protons in 0\%--5\% central Pb + Pb collisions at LHC energies of \(\sqrt {s_{NN}}\) =  2.76, 3.94, and 5.52 TeV. The theoretical distributions are shown for  
\(\lambda =\) 0.2.} 
\end{center}
\end{figure}
\begin{figure}
\begin{center}
\includegraphics[width=9cm]{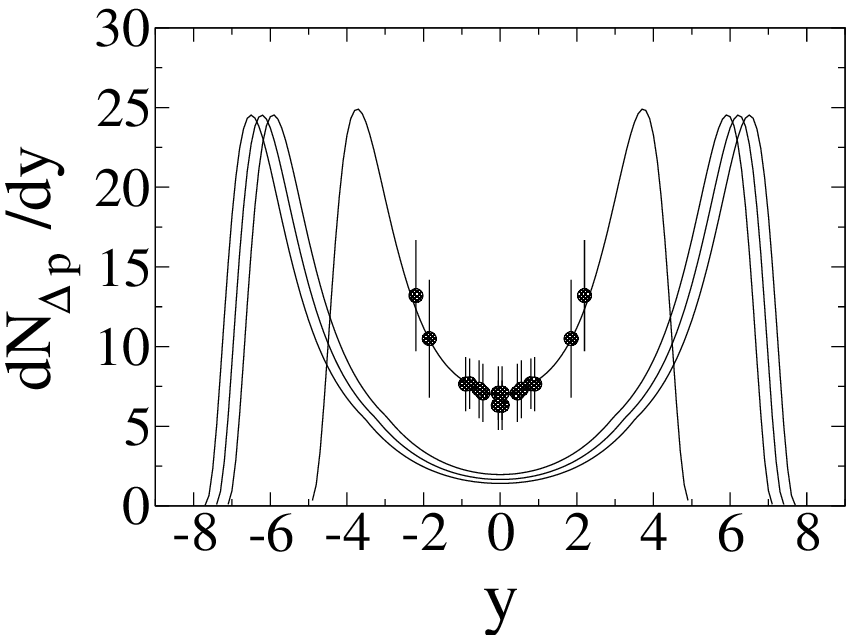}
\caption{\label{fig4}Calculated rapidity distributions of net protons in 0\%--5\% central Pb + Pb collisions at LHC energies of \(\sqrt {s_{NN}}\) = 2.76, 3.94, 5.52 TeV. Our result for central Au + Au collisions at RHIC energies of 0.2 TeV is compared with BRAHMS data \cite{bea04} in a  \(\chi^2-\) minimization
as in Fig.~\ref{fig1}.}
\end{center}
\end{figure}

In Fig.~\ref{fig3} we display the energy dependence of our net-proton central Pb + Pb results near midrapidity for \(\sqrt {s_{NN}}\) = 2.76, 3.94, and 5.52 TeV. At \(y = 0\) the corresponding values of 
\(dN/dy\) are 1.9, 1.7, and 1.4.

A description for the net-proton rapidity distribution within the relativistic diffusion model (which is not based on QCD, but on nonequilibrium-statistical physics) had been developed in \cite{wol07}. There the predicted midrapidity value for central Pb + Pb at the LHC energy of \(\sqrt {s_{NN}}\) =  5.52 TeV is 
\(dN/dy \simeq\) 1 -- 2.5 depending on the model parameters and hence, comparable to the present QCD-based result.

To emphasize how our midrapidity results are embedded into the overall shape of the rapidity distribution for net protons (baryons) in central relativistic Pb + Pb  collisions at LHC energies,
we show the total rapidity density distribution functions for
the BRAHMS Au + Au data at 0.2 TeV and for the three LHC energies
\(\sqrt {s_{NN}}\) = 2.76, 3.94, 5.52 TeV in Fig.~\ref{fig4}.  

Here we have used for the mid-rapidity valley Eq.~(\ref{eq:valley}) (as in Figs.~1, 2) matched at the point \(x_2=0.01\) with the parametrization (cf. Eqs.~(7) and (8) in \cite{mrst01} ) of the  valence quark distribution function which describes the larger rapidities.
As is evident from the figure, the transition between the two regimes is fairly smooth. Both up- and down-quark parton distribution functions are considered with the appropriate weights.

To conclude, we have presented predictions for net-proton and net-kaon rapidity distributions in central Pb + Pb collisions at LHC energies, with an emphasis on the midrapidity region where data will be available in the near future. We have set up and used a transparent QCD-based model and well-established parton distribution functions in the context of saturation physics, and we expect that
our predictions at midrapidity will turn out to be reliable.

We have shown that hadronization does not influence the slope of the net-hadron rapidity distributions
since net-proton and net-kaon rapidity distributions are related through a constant factor. Hence, net-baryon and net-charge transport provide a powerful tool to investigate initial-state dynamics in heavy-ion collisions. Finally, we have extracted a value for the strong-coupling constant \(\alpha_s \approx 0.2\) both from net-proton and net-kaon rapidity distributions in Au + Au at RHIC energies.\\






This work has been supported
by 
the ExtreMe Matter Institute EMMI.

\vspace{1cm}

\end{document}